\documentstyle[prc,aps,psfig,preprint,tighten]{revtex}
\begin{document}
\draft
\title{On the treatment of confinement in three-quark calculations}
\author{Z. Papp \\
Institute of Nuclear Research of the 
Hungarian Academy of Sciences, \\
P.O. Box 51, H--4001 Debrecen, Hungary}
\date{\today}
\maketitle

\begin{abstract}
The treatment of confining interactions in non-relativistic three-quark
systems is revised. Usually in the Faddeev equations 
the Faddeev components are coupled by the total potential. In the new 
treatment the Faddeev components are coupled only by the non-confining 
short-range part of the potential, allowing thus its channel-by-channel
investigation. 
The convergence in angular momentum channels is much faster.
\end{abstract}

\pacs{PACS number(s): 12.39.Jh; 14.20,-c}

\vspace{0.5cm}

The Faddeev equations are the fundamental equations 
of three-body problems.  Besides giving a unified
formulation, they are superior to the
Schr\"odinger equation in many respects:
in incorporating boundary conditions,
in treating symmetries, in handling correlations, etc., and only 
this formalism can handle all types of interactions.
These unique advantages also appear in
three-quark calculations  \cite{s-brac,gpp,frasc}.
It has been observed, however, that in the Faddeev scheme
many angular momentum channels are needed. So, in practice, the
apparent advantages of the Faddeev method are
burdened by the need for many angular momentum channels.

A similar situation occurs in atomic physics. There are several 
bound-state atomic three-body calculations using Faddeev technique.
They demonstrate the power of this  scheme in all kinds of mass ratios
and also for excited states, on the expense of many angular momentum 
channels, however. 
In a recent publication, Ref.\ \cite{pzatom}, I have pointed out 
that this phenomenon is the consequence of the inadequate way of using
the Faddeev technique. A new scheme has been proposed, which means 
splitting the Coulomb interactions into long-range and short-range terms,
and applying the Faddeev decomposition only for the short-range terms.
This way all the asymptotically important terms, like the kinetic energy
and the long-range part of the interaction were treated  on equal 
footing and in an asymptotically exact way. Numerical studies showed 
that in this scheme,
to reach a good accuracy, much less angular momentum channels are needed.

The role of the Coulomb and confinement potentials in three-body atomic 
and in three-quark systems is similar: both kinds of potentials modify 
the character of the asymptotic motion, and thus, following the concept
of scattering theory, both of them
should be treated on equal footing with the kinetic energy.
In the present work, taking over the concept of Ref.\ \cite{pzatom}, 
I propose a new way of treating confinement potentials in three-quark
calculations. I split the quark-quark interaction into a long-range 
confining and a short-range non-confining interactions, 
and apply the Faddeev decomposition only for the 
non-confining part. In the resulting Faddeev equations, in complete 
agreement with our physical expectation, the long-range confining parts
play a similar role as the kinetic energy, thus the asymptotics of the
Faddeev components are determined by
them together. The power of the new method is demonstrated by numerical
illustrations.

The Sch\"odinger equation of a three-quark system reads 
\begin{equation}
(H^0 + v_\alpha+ v_\beta + v_\gamma)|\Psi \rangle = E |\Psi \rangle,
\label{H}
\end{equation}
where $H^0$ is the three-body kinetic energy 
operator and $v_\alpha$ denotes the 
quark-quark interaction in subsystem $\alpha$. 
In the conventional method \cite{s-brac}, the Faddeev procedure
is applied literally.
The  the wave function $\Psi$ is written as a superposition
of three Faddeev components
\begin{equation}
|\Psi \rangle
=|\psi_{\alpha} \rangle+|\psi_{\beta} \rangle+|\psi_{\gamma} \rangle,
\end{equation}
and the components are required to satisfy
the set of Faddeev equations
\begin{equation}
 (E- H^0 -   v_\alpha )   |\psi_{\alpha} \rangle=  v_\alpha \left[
|\psi_{\beta}\rangle +  | \psi_{\gamma}\rangle \right],
\label{feqs}
\end{equation}
with a cyclic permutation for $\alpha, \beta, \gamma$.
Of course, the sum of three Faddeev equations gives back the original
Schr\"odinger equation (\ref{H}).

Since $v_\alpha$ depends only on $\xi_\alpha$ Jacobi distance,  
it is natural to express the Faddeev components with the corresponding
$(\vec{\xi}_\alpha,\vec{\eta}_\alpha)$ Jacobi coordinates. Each Faddeev 
component is expanded on an angular momentum basis, which should
carry also the necessary spin and isospin indices. 
Usually, the energy reference corresponds to the case where all particles 
are infinitely separated. For confining particles this makes no sense.
Since the asymptotics of $|\psi_{\alpha} \rangle$ in the $\eta_\alpha$
coordinate is determined by the kinetic energy component 
$h^0_{\eta_\alpha}$
only, this asymptotics is strongly dependent on the energy reference.
The practical way out of this problem is choosing the energy reference 
in such a way that the relevant eigenvalues appear as very deeply 
bound states. This way the asymptotics of  $|\psi_{\alpha} \rangle$ 
in the $\eta_\alpha$ coordinate becomes similar to a confinement 
asymptotics. In Ref.\ \cite{s-brac}
the equations were solved in configuration space. In Ref.\ \cite{gpp}
a simplified version of the method of Ref.\ \cite{pzwp} was used.
In fact, the Faddeev components were expanded in terms of 
Coulomb--Sturmian functions.

It should be noted that the splitting of the wave function $|\Psi \rangle$
into Faddeev components is not unique. If the potential
$v_\alpha $ is split into a confining and non-confining terms,
\begin{equation}
v_\alpha =v_\alpha^c + v_\alpha^{nc},
\label{split}
\end{equation}
where $v_\alpha^c$ should contain all the asymptotically relevant terms,
like constant and confinement terms, and $v_\alpha^{nc}$ is short-ranged
compared to $v_\alpha^c$,
the Faddeev procedure can also be applied only for the non-confining 
short-range part.
The  wave function $|\Psi \rangle$ can also be written as a superposition
of the modified Faddeev components,
\begin{equation}
|\Psi \rangle
=|\widetilde{\psi}_{\alpha} \rangle+
|\widetilde{\psi}_{\beta} \rangle+|\widetilde{\psi}_{\gamma} \rangle,
\label{psitilde}
\end{equation}
which are required to satisfy
the set of modified Faddeev equations
\begin{equation}
 (E- H^0 - v_\alpha^c - v_\beta^c - v_\gamma^c -v_\alpha^{nc} )   
 |\widetilde{\psi}_{\alpha} \rangle=  v_\alpha^{nc} [
|\widetilde{\psi}_{\beta}\rangle +  |\widetilde{\psi}_{\gamma}\rangle ],
\label{feqm}
\end{equation}
with a cyclic permutation for $\alpha, \beta, \gamma$.
Of course, the sum of three Faddeev equations also gives back the original
Schr\"odinger equation (\ref{H}).
In this equation the  asymptotics of $|\widetilde{\psi}_{\alpha}\rangle $ 
along the coordinate $\eta_\alpha$ is determined not only by the kinetic 
energy, but also by the sum of confining interactions 
$v_\beta^c + v_\gamma^c$,  and thus the asymptotics does not 
depend on the choice of energy reference.

The Faddeev procedure is not merely rewriting
the wave function $|\Psi \rangle$ as a sum of three components: the
decomposition should also act as an asymptotic filtering \cite{vanzani}.
While the wave function $|\Psi \rangle$ describes all the three different 
two-cluster fragmentations each Faddeev 
component describes only one fragmentation.
The necessary condition for that is that  the term
$v_\beta^c + v_\gamma^c$ on the left hand side of Eq.\ (\ref{feqm})
should not generate bound states. Since they are
infinite range confining interactions this is not possible. However, 
if the splitting (\ref{split}) is performed in such a way that $v^c$ 
contains a repulsive core the bound states generated by  
$v_\beta^c + v_\gamma^c$ can be removed from the spectrum of physical 
interest. So, in the energy region of physical interest the 
decomposition (\ref{psitilde}) acts as an asymptotic filtering as well.

Eqs.\ (\ref{feqm}) can be solved as before. For comparison, I present here
the convergence of the energy (mass) of barions by increasing the 
number of angular momentum channels using both Eqs.\ (\ref{feqs}) 
and Eqs.\ (\ref{feqm}). In Refs.\ \cite{gpp} and \cite{frasc} the 
barions are described as three-quark systems interacting by two different
kind of parametrization of a linear confinement plus 
Goldstone-boson-exchange potential,
\begin{equation}
V_{qq}(\vec{r})=V_0 + C r+V_\chi .
\end{equation}
In the parametrization of Ref.\ \cite{gpp}
the strength of the confining force is relatively week,
$C=0.474\;\mbox{fm}^{-2}$, while in the other
one, in the alternative parametrization of Ref.\ \cite{frasc}, 
it is rather strong, $C=0.77\;\mbox{fm}^{-2}$. 
The following kind of splitting of the 
total potential is adopted: 
\begin{equation}
v^c=V_0+C r+ V_g \exp(-(r/r_g)^2) 
\end{equation}
 and 
\begin{equation}
v^{nc}=V_\chi-V_g \exp(-(r/r_g)^2),
\end{equation}
with $r_g=1.0\;\mbox{fm}$ and $V_g=3.0\;\mbox{fm}^{-1}$ in both cases.
 The role of the Gaussian term in $v^c$ is 
to remove the spurious states from the spectrum, and, in the other hand, 
it should not destroy the convergence with respect to angular momentum 
channels. The results are given in Table I. The convergence
in the new scheme, Eqs.\ (\ref{feqm}), is much faster;
in fact channels up to $l=2$ are completely sufficient in both cases. 
The converged results from Egs.\ (\ref{feqm}) are in complete agreements
with the fully converged results from Eqs.\ (\ref{feqs}), which
in Ref.\  \cite{frasc} were cross-checked with the stochastic variational
method \cite{kalman}.

Summarizing, in this paper I have proposed a new way of writing down
the Faddeev equations for confining potentials. In this new scheme all the 
asymptotically important parts, like constant and confining parts, are
treated on equal footing with the kinetic energy, and, while all the 
previous advantages of the Faddeev technique are preserved, 
the convergence with respect to angular momentum channels is much faster. 
The fact that the confining and non-confining parts are treated in 
different ways makes the channel-by-channel investigation of the 
physically interesting non-confining part possible. For example from 
Table I it is obvious, that in the model of Refs.\ \cite{gpp} and 
\cite{frasc} the higher partial waves of the non-confining interaction 
do not contribute to the mass of the nucleon and its excitations. 
This conclusion could have
hardly been inferred from the results of Eqs.\ (\ref{feqs}),
and even less from other non-Faddeev calculations.
I believe that for analyzing the quark-quark interaction in 
non-relativistic quark model by investigating
the properties of barions the modified Faddeev equations (\ref{feqm})
provide the best tools.

This work has been supported by OTKA under Contracts No. T17298 and
T026233.

{}

\begin{table}[tbp]
\caption{Convergence of the mass
of the nucleon and ``excited'' nucleons
 with respect to angular momentum channels
taken into account  up to
$l=0$, $l=1$, $l=2$, $l=3$ and $l=4$.
The energy values are given in MeV.
}
\label{tabconv}
\begin{tabular}{rccccc}
& \multicolumn{5}{c}{Angular momentum channels}  \\ 
 & $l=0$ & $l=1$ & $l=2$ & $l=3$ & $l=4$ \\ \hline 
\multicolumn{6}{c}{Parametrization of Ref.\ \cite{gpp} }  \\ \hline 
$N$ with Eqs.\ (\ref{feqs}) & 950 & 945 & 939  &  939
&  939 \\  
$N$ with Eqs.\ (\ref{feqm}) & 939 & 939 & 938  &  938 &  938 \\  \hline
$N^{\star}$ with Eqs.\ (\ref{feqs}) & 1574 & 1565 & 1510 & 
1502 &  1493 \\  
$N^{\star}$ with Eqs.\ (\ref{feqm}) & 1495 & 1490 & 1490 & 
1490 &  1490 \\  \hline
$N^{\star \star}$ with Eqs.\ (\ref{feqs}) & 1859 & 1780 & 1724 & 
1698 & 1690  \\ 
$N^{\star \star}$ with Eqs.\ (\ref{feqm}) & 1704 & 1689 & 1681 & 
1682 & 1681 \\
 \hline
\multicolumn{6}{c}{Alternative parametrization of Ref.\ \cite{frasc} }  
\\ \hline 
$N$ with Eqs.\ (\ref{feqs})         & 1050 & 1026 & 970  & 
 965 &  959 \\  
$N$ with Eqs.\ (\ref{feqm})         & 953  & 953  & 942  & 
 942 &  942 \\  
\hline
$N^{\star}$ with Eqs.\ (\ref{feqs}) & 1566 & 1551 & 1508 &  
1498 & 1482 \\  
$N^{\star}$ with Eqs.\ (\ref{feqm}) & 1478 & 1471 & 1468 &  
1468 & 1468 \\  
\hline
$N^{\star \star}$ with Eqs.\ (\ref{feqs}) & 1900 & 1802 & 1780 & 
1756 & 1748 \\ 
$N^{\star \star}$ with Eqs.\ (\ref{feqm}) & 1780 & 1723 & 1714 & 
1714 & 1714 
\end{tabular}
\end{table}

\end{document}